\documentclass[amsmath,amssymb]{article}

\usepackage{graphicx}
\usepackage{dcolumn} 
\usepackage{bm, bbm}      
\usepackage{curves}
\usepackage{epic}
\usepackage{wasysym}
\usepackage{epsf, times}
\usepackage{subfigure}
\usepackage{amsthm}

\begin{document}


\title{
Coulomb physics in spin ice \\
from magnetic monopoles to magnetic currents
      }

\author{
Claudio Castelnovo \\ 
Rudolf Peierls Centre for Theoretical Physics and Worcester College \\ 
University of Oxford \\ 
Oxford, OX1 3NP, United Kingdom
       }

\maketitle
%
%

In the second half of the past century it became apparent 
that the low temperature behaviour of condensed matter systems can often be 
described by modeling their excitations as \emph{quasiparticles} immersed in 
an \emph{effective vacuum}, whose properties derive directly from those of 
the low temperature phase of the system. 
Whereas in the search for the fundamental constituents in our universe 
we are bound to look for what is already there, 
the combinatorial nature of the periodic table could then be harvested to 
realise an endless variety of new vacua, with relative exotic excitations. 
For instance, particles that carry a fraction of the electronic charge were 
found in polyacetylene and fractional quantum Hall systems; or 
electron-like quasiparticles that carry charge but no spin 
(spin-charge separation) in SrCuO$_2$. 

In late 2007, it was suggested that the unconventional low temperature 
behaviour of a class of rare earth titanates (namely, Dy$_2$Ti$_2$O$_7$ and 
Ho$_2$Ti$_2$O$_7$), dubbed spin ice, can be understood in terms of point-like 
quasiparticle excitations, with the exceptional property of carrying a 
\emph{net magnetic charge}!~\cite{Castelnovo2008} 

These materials are localised spin systems where the magnetic degrees of 
freedom (the rare earth ions) form a corner sharing tetrahedral lattice, 
with two distinctive features: 
(i) a strong single ion anysotropy causes the spins to be uniaxial, with an 
energy barrier in excess of $100$~K; and 
(ii) the interactions between the large rare earth spins are dominated by 
the magnetic dipolar coupling. 

\begin{figure}[ht]
\begin{center}
\includegraphics[height=0.3\columnwidth]{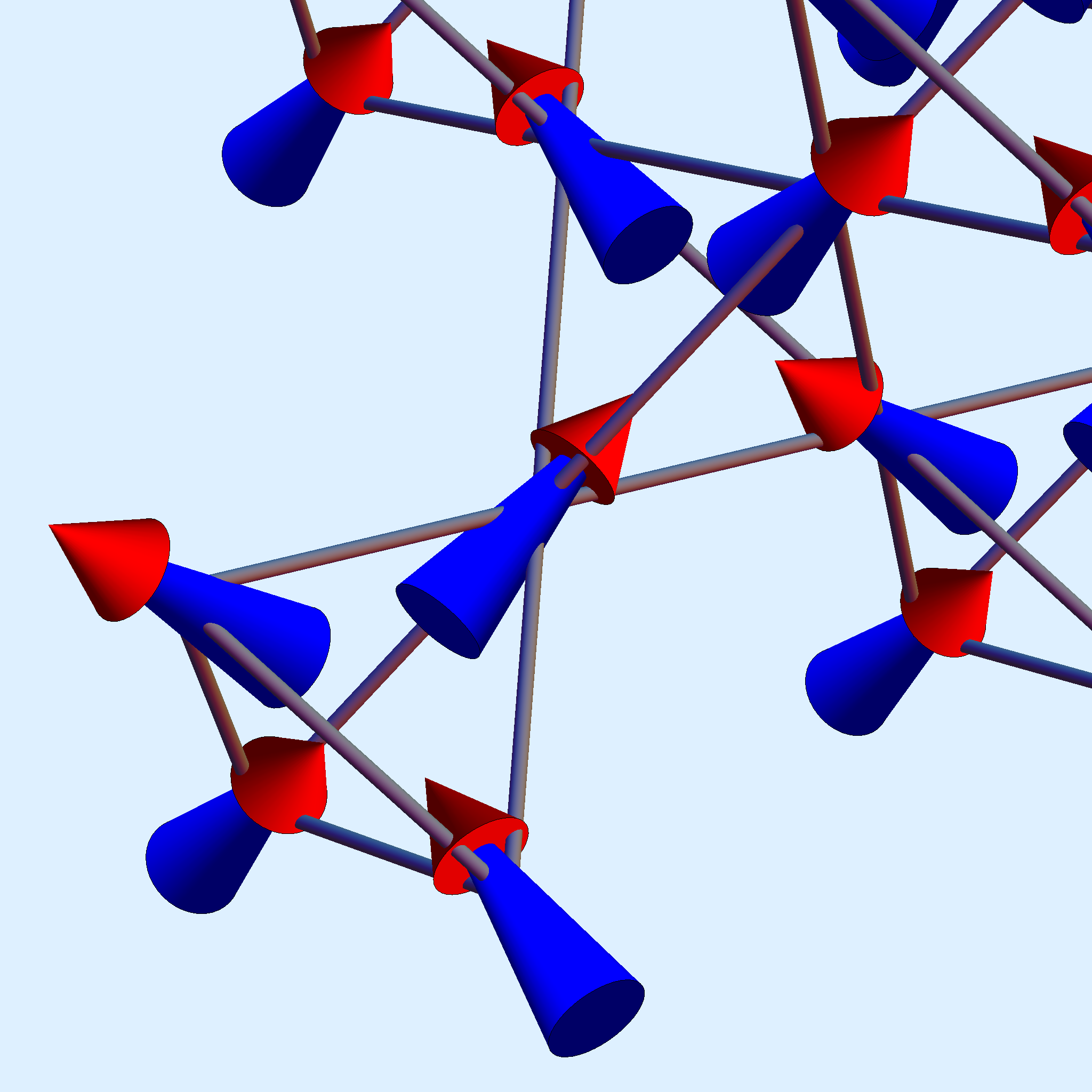}
\hspace{0.2 cm}
\includegraphics[height=0.3\columnwidth]{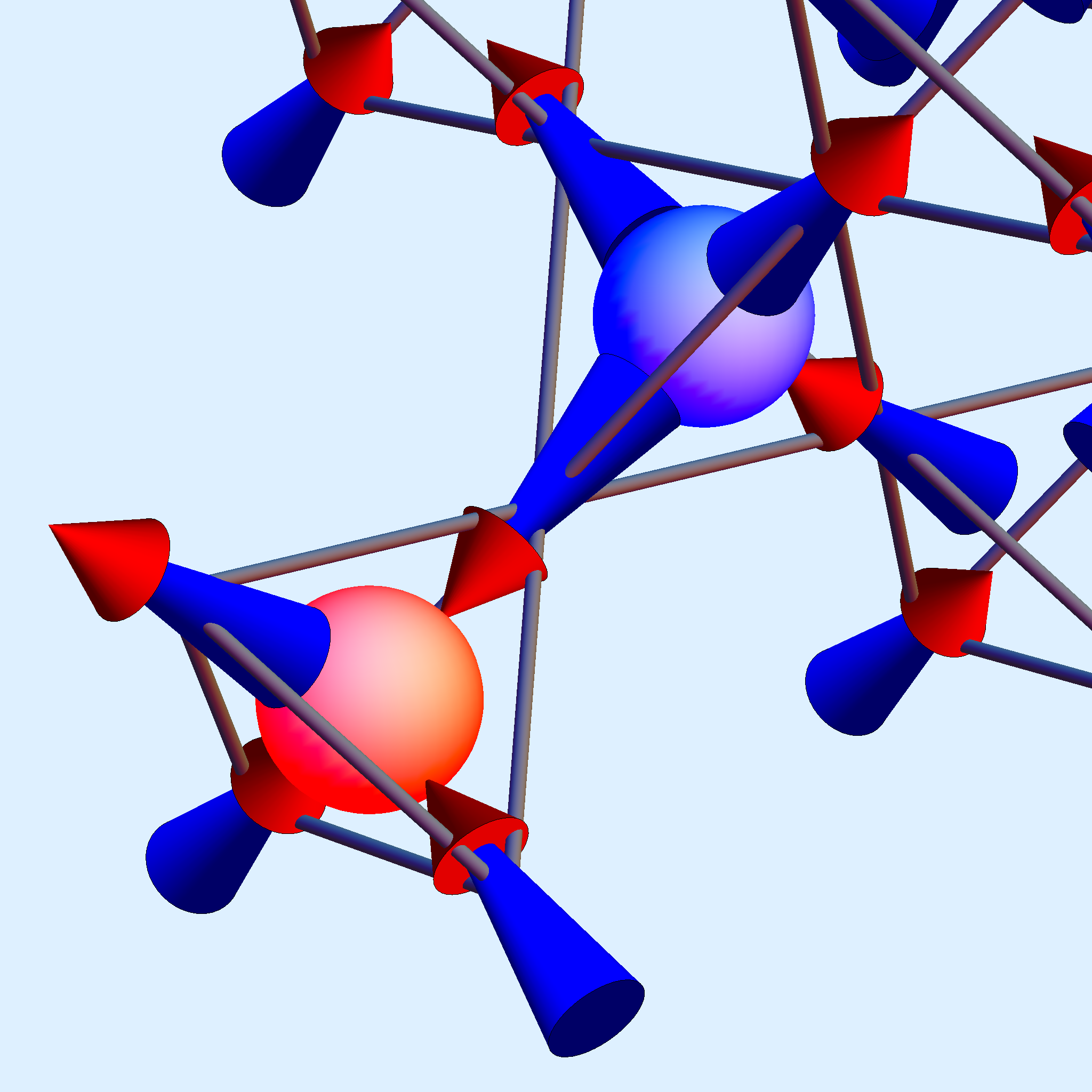}
\hspace{0.2 cm}
\includegraphics[height=0.3\columnwidth]{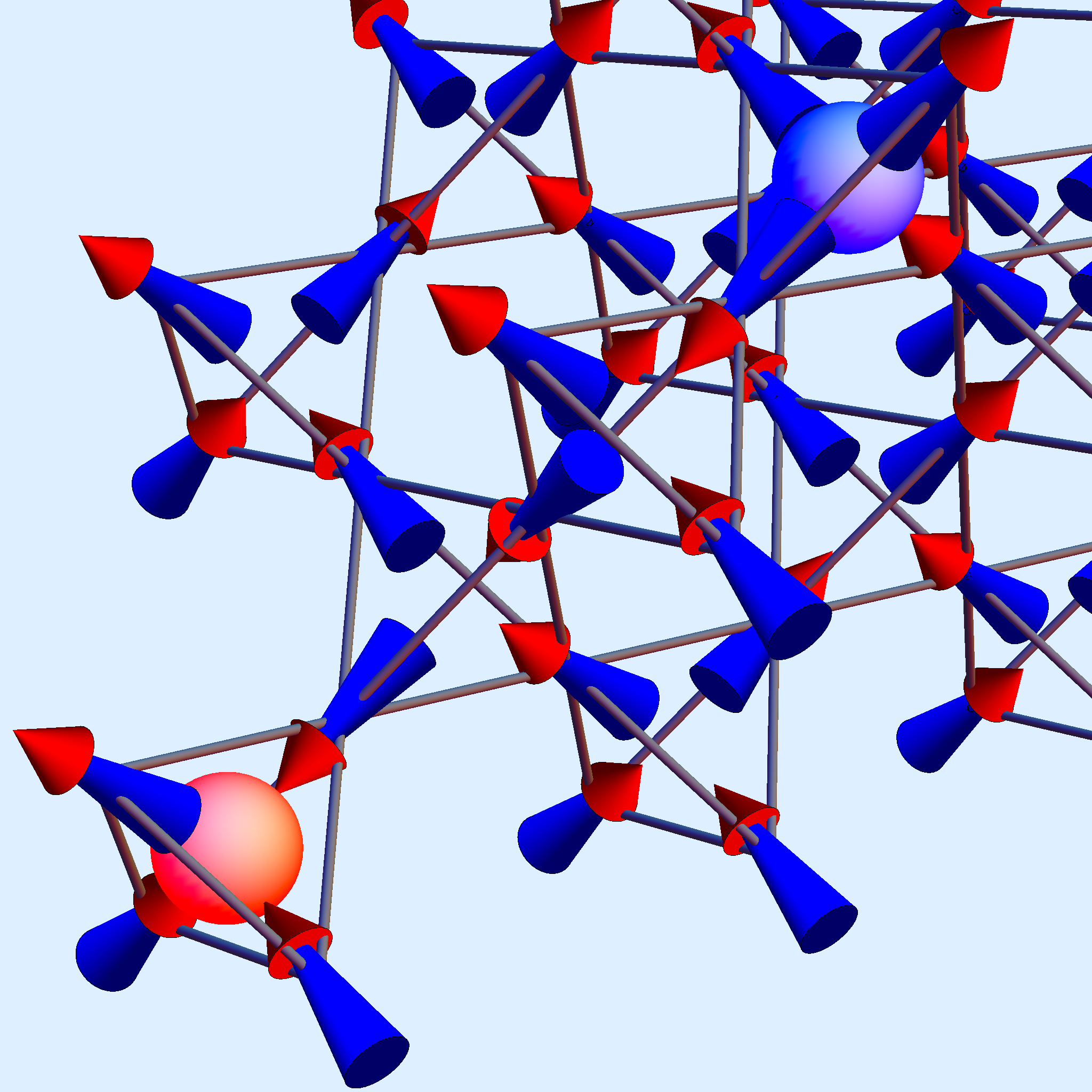}
\end{center}
\caption{
\label{fig: monopoles} 
Pictorial representation of a spin configuration in rare earth titanates. 
Left panel: 
configuration that satisfies the two-in, two-out ice rules, whereby 
each tetrahedron has a vanishing net magnetic charge. 
From the point of view of the Wien effect, this is analogous to undissociated 
H$_2$O molecules in water. 
Middle panel: 
the reversal of a spin introduces two adjacent defective tetrahedra 
(red and blue sphere), which in the same analogy correspond to a bound 
[H$_3$O$^+$ OH$^-$] pair. 
Right panel: 
an appropriate rearrangement of spins can lead to the separation of two 
defects without introducing other violations of the ice rules. 
If the distance between the defects becomes larger than the screening length, 
they dissociate and behave as free charged ions in water. 
}
\end{figure}

The combination of these properties results in a \emph{frustrated ferromagnet}, 
whose low-energy configurations are characterised by spin-spin correlations 
akin to the proton-proton correlations in 
water ice -- hence the name 
spin-ice. Every tetrahedron satisfies the so-called ice rules: two 
spins point in and two point out (see Fig.~\ref{fig: monopoles}, left panel). 
An exponentially large number of spin configurations satisfy these rules, 
leading to an extensively degenerate low-temperature phase, customary of 
frustrated systems. 

The peculiar structure of the low-temperature phase in spin ice manifests 
itself in the nature of its excitations. 
An excited state is generated when the orientation of a spin is reversed 
with respect to its lowest energy state, and the two adjacent tetrahedra no 
longer fullfill the ice rules (see Fig.~\ref{fig: monopoles}, middle panel). 
Defective tetrahedra can then be separated by means of spin rearrangements 
that do not introduce further violations of the ice rules 
(Fig.~\ref{fig: monopoles}, right panel), at a free energy cost inversely 
proportional to the distance between the defects. 
The barrier to separate two adjacent defects to infinity is therefore finite 
(they are deconfined), and a \emph{single} defective tetrahedron 
represents an \emph{elementary excitation} in spin ice 
-- a rare example of \emph{fractionalisation} in three dimensions. 

This phenomenon is even more striking if we consider the nature of these 
defects. Spins carry magnetic moment, and the ice rules imply that 
the spins are oriented with two north poles and two south poles close to 
the centre of each tetrahedron, in a locally neutral arrangement. 
Defective tetrahedra with three spins pointing in and one pointing out 
(or vice versa) are local excesses of north (south) poles. 
The elementary excitations in spin ice are therefore fractions of the 
underlying spin degrees of freedom, in that they carry a net magnetic 
charge -- the closest (classical) realisation of a magnetic monopole to date! 

The presence of magnetic monopoles allows to explain the 
liquid-gas structure of the experimental phase diagram in a magnetic 
field~\cite{Castelnovo2008} -- a feature that was 
reported as ``unprecedented in a localised spin system'' -- as well as 
an exceptional increase in the characteristic time scales for magnetic 
relaxation at low temperatures.~\cite{Jaubert2009} 

Since the theoretical proposal, a broad experimental effort materialised at 
lightning speed to find more direct confirmations of the 
existence of these monopoles. 
Three of the major research groups in the field, based in 
Germany,~\cite{Morris2009} 
England,~\cite{Fennell2009} 
and Japan,~\cite{Kadowaki2009} 
succeeded. 
Using neutron scattering techniques, distinctive evidence was found for both 
the characteristic reversed spin chains separating pairs of monopoles, 
and for the dipolar correlations between the underlying spins, fathering the 
peculiar nature of these excitations. 
Together with susceptibility and heat capacity measurements, the experimental 
results strongly support the idea that the low temperature behaviour of spin 
ice is akin to that of a gas of free magnetically charged particles, 
i.e., a \emph{magnetic Coulomb liquid}. 

What new phenomena can one expect in materials where the 
low energy phase exhibits magnetic monopole excitations? 

To begin with, the onset of the ice rules ought 
to be substantially different from phase 
ordering kinetics in more conventional magnets. 
Concepts like domain growth and coarsening are replaced by diffusion and 
annihilation of Coulomb-interacting point-like defects.~\cite{Castelnovo2009} 

O.~Tchernyshyov, writing in Nature,~\cite{Tchernyshyov2008} further speculated 
that ``learning how to move magnetic monopoles around would be a step towards 
technologies such as magnetic analogues of electric circuits and magnetic 
memories operating on the atomic scale''. 

One important difference between spin ice monopoles and free magnetic charges 
is that a steady flow (direct current) is forbidden: 
a monopole moving throught the lattice orients the spins along its path in 
a way that does not allow another monopole with the same charge to follow the 
same path. 
However, there are no reasons of principle that prevent alternating currents. 
A concrete step in this direction was cleverly accomplished by Steve 
Bramwell and collaborators, combining a 1934 theory by Onsager on the 
behaviour of electrolytes, with state of the art muon spin rotation 
measurements.~\cite{Bramwell2009} 

Weak electrolytes are known to exhibit a non-linear increase in dissociation 
constant $K$ in presence of an applied electric field -- known as the second 
Wien effect. 
Consider for simplicity the familiar case of autoionisation in water. 
While most of the molecules have no net charge, a small fraction 
of them is dissociated into H$_3$O$^+$ and OH$^-$ ions. 
Opposite ions attract each other via Coulomb interactions, and 
free charges appear only at the cost of overcoming the 
Coulomb energy barrier to separate them beyond the screening length. 
The system is therefore governed by two successive thermal equilibria, 
\begin{equation}
2{\rm H}_2{\rm O} 
\;\rightleftharpoons\; 
\left[ \, {\rm H}_3{\rm O}^+ \: {\rm OH}^- \, \right] 
\;\rightleftharpoons\; 
{\rm H}_3{\rm O}^+ \, + \, {\rm OH}^- 
. 
\label{eq: dissociation equilibria}
\end{equation}
%
%
An applied electric field $E$ reduces the barrier for bound pairs to become 
free charges, which affects the dissociation constant $K$ of the second 
process in Eq.~(\ref{eq: dissociation equilibria}). 
The central result in Onsager's theory quantifies this change perturbatively 
in the applied field strength,~\cite{Onsager1934} 
\begin{eqnarray}
K(E) &=& K(0) \left[ 1 + b + \frac{b^2}{3} + \ldots \right] 
\label{eq: Onsager (I)}
\\ 
b &=& \frac{e^3 E}{8\pi\varepsilon_0 \, k_B^2 T^2}
, 
\label{eq: Onsager (II)}
\end{eqnarray}
where $e$ is the ionic charge, $\varepsilon_0$ the 
electric permittivity of the vacuum, $k_B$ Boltzmann's constant, and $T$ the 
temperature. 
A remarkable feature of this thermodynamic result is that it allows 
to determine experimentally the value $e$ of the carrier charge. 

After a sudden change in the applied field, the dissociation constant $K$ 
relaxes to its equilibrium value exponentially, with a decay rate $\nu_K$ 
proportional to the conductivity of the system. 
Onsager showed that in the limit of small free charge density, 
the conductivity is in turn proportional to the square root of $K$, so that 
\begin{equation}
\frac{\nu_K(E)}{\nu_K(0)} 
= 
\sqrt{\frac{K(E)}{K(0)}} 
\simeq 
1 + \frac{b}{2} 
. 
\label{eq: diss const decay}
\end{equation}

Onsager theory successfully applies to several electrolytes, solid or 
liquid. If the low-temperature description of spin ice in terms of 
magnetic charges is correct, then the theory should bear 
relevance in that context as well, provided we replace 
$e$ with the magnetic monopole charge $Q$, 
$E$ with the applied magnetic field $B$, and 
$\varepsilon_0$ with the permeability of the vacuum $\mu_0$. 
There is however an important difference: the motion of monopoles under 
the influence of the field leads to a change in the magnetisation of the 
underlying spin configuration. 
In the weak field limit, Bramwell and co-workers argue that the magnetisation 
change per unit forward reaction is constant, independent of the magnetic 
field, and therefore the relaxation rate $\nu_\mu$ of the magnetisation decay 
after a field quench is proportional to $\nu_K$, 
\begin{eqnarray}
\frac{\nu_\mu(B)}{\nu_\mu(0)} 
&=& 
\frac{\nu_K(B)}{\nu_K(0)} 
= 
\sqrt{\frac{K(B)}{K(0)}} 
\simeq 
1 + \frac{b}{2} 
\label{eq: mag decay 1}
\\ 
b &=& \frac{\mu_0 \, Q^3 B}{8\pi \, k_B^2 T^2}
. 
\label{eq: mag decay 2}
\end{eqnarray}
Eq.~(\ref{eq: mag decay 2}) allows then to obtain the value of $Q$ 
from relative experimental measurements sensitive to the relaxation of the 
magnetisation of the sample. 

\begin{figure}[ht]
\begin{center}
\includegraphics[height=0.5\columnwidth]{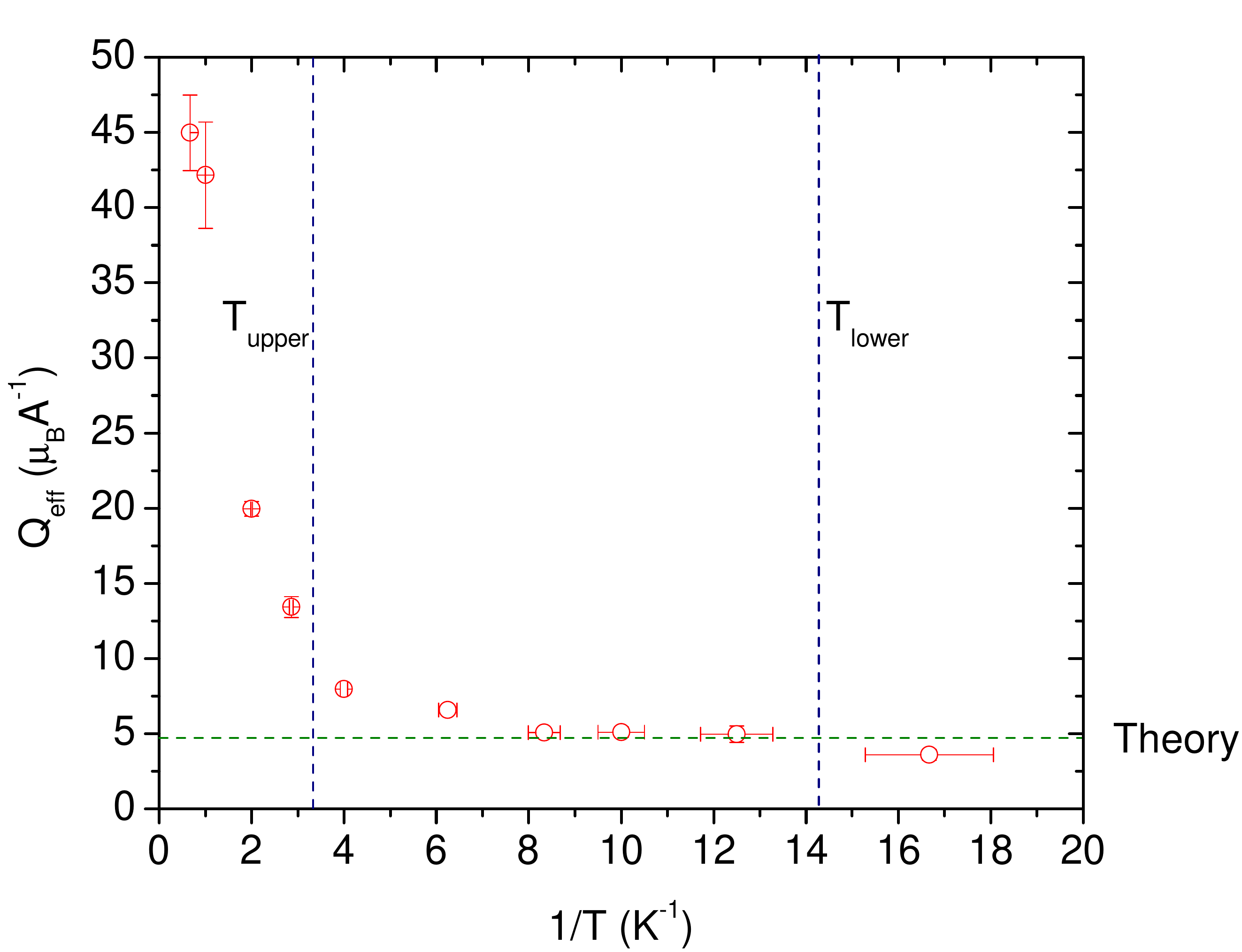}
\end{center}
\caption{
\label{fig: monopole charge} 
Experimental values of the magnetic carrier charge in spin 
ice,~\cite{Bramwell2009} compared to the 
theoretical prediction (horizontal line).~\cite{Castelnovo2008} 
}
\end{figure}

The experimental probe of choice to this purpose is transverse field 
muon spin rotation. The spins of muons implanted in a sample 
undergo a characteristic oscillatory relaxation behaviour as they 
precess about the applied field, subject to dephasing caused by 
the fluctuating local field due to the magnetisation of the sample. 
At low temperature, when the fluctuations in magnetisation are sufficiently 
slow, dephasing leads to an exponential decay envelope for the oscillatory 
behaviour, with a characteristic decay rate proportional to $\nu_\mu$. 
The results by Bramwell and co-workers on Dy$_2$Ti$_2$O$_7$ show 
clear evidence of the scaling behaviour 
in Eqs.~(\ref{eq: mag decay 1}-\ref{eq: mag decay 2}) in the temperature range 
$0.07 < T < 0.3$~K,~\cite{Bramwell2009} and the measured value of the 
magnetic charge $Q \sim 5$~$\mu_B$\AA$^{-1}$ 
(see Fig.~\ref{fig: monopole charge}) is in good agreement with 
the one predicted by the theory.~\cite{Castelnovo2008} 

Not only do these measurements provide further compelling evidence of the 
presence of magnetic monopole excitations in spin ice materials, and of their 
magnetic Coulomb interactions. 
They also show that spin ice monopoles respond to external magnetic fields 
(to leading order) in the same way as electric charges do for instance 
in water, making Dy$_2$Ti$_2$O$_7$ the first material of 
a class that one might rightfully call \emph{magnetolytes}. 

What Bramwell and co-workers have accomplished is the first step towards 
determining whether macroscopic alternating currents are ultimately 
achievable in spin ice. 
This gives new emphasis to the study of magnetic charges in condensed 
matter system, as well as a concrete perspective to potential technological 
applications. 
Further experimental and theoretical efforts are needed to fill the gap 
from magnetolytes to \emph{magnetricity} -- 
time will tell how far these monopoles can travel. 
%
%

\section*{Acknowledgments}

Financial support from the EPSRC Postdoctoral Research Fellowship EP/G049394/1 
is gratefully acknowledged. 
%
%

%
%

\section*{Table of Contents}

\includegraphics[height=0.1\columnwidth]{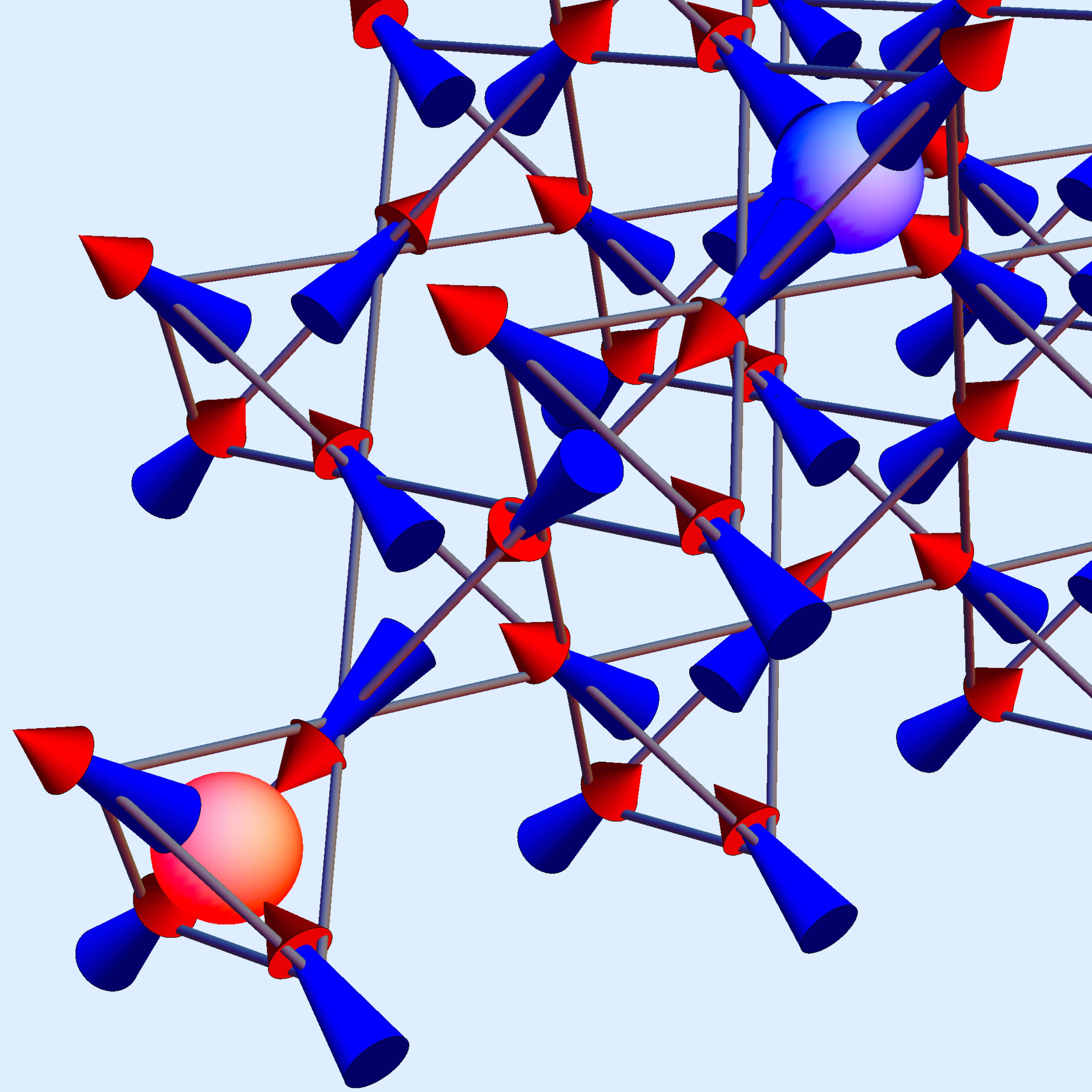}
\hspace{0.3 cm}
\begin{minipage}{10.5 cm}

\vspace{-0.4 cm}
It was proposed that spin ice, a class of rare earth titanates, 
hosts magnetic monopoles as elementary excitations. 
Recently, Bramwell and co-workers measured 
the Wien effect in Dy$_2$Ti$_2$O$_7$, 
directly probing the nature of these monopoles 
and making Dysprosium titanate the first example of a \emph{magnetolyte}. 
\end{minipage}
%
%

\end{document}